# Quantum Monte Carlo study of magnetism and chiral d+id-wave superconductivity in twisted bilayer graphene[*]

FANG Shichao, LIAO Xinyi

School of Mathematics and Physics, Nanyang Institute of Technology, Nanyang 473004, China

**Abstract**

We employ a large-scale, unbiased constrained-path quantum Monte Carlo method to systematically simulate the effective two-orbital Hubbard model for twisted bilayer graphene in order to gain deeper insight into the relationship between correlated states and the superconducting pairing mechanism in twisted bilayer graphene, as well as the influence of the twist angle on superconductivity. Initially, we investigate the modulation of superconductivity by nearest-neighbor attractive Coulomb interactions, demonstrating that electron-phonon coupling plays a significant role in the system. Our numerical results reveal that the superconducting state is dominated by chiral NN-d+id superconducting electron pairing symmetry, and that such nearest-neighbor attractive Coulomb interactions significantly enhance the effective long-range pairing correlation function of chiral NN-d+id wave. From this perspective, it is evident that the electron-phonon coupling positively contributes to the superconductivity of the system.Then, we explore how the twist angle affects the superconducting state. The flat-band structure caused by hopping anisotropy reflects the different twist angles of the system. Our results show that as the twist angle deviates downward from 1.08°, the effective pairing correlation function of the chiral NN-d+id wave increases substantially. Conversely, as the twist angle exceeds 1.08°, the effective correlation function of the chiral NN-d+id wave exhibits a tendency of decline. These results suggest that further reduction of the twist angle may lead to higher superconducting transition temperature in twisted bilayer graphene system.Finally, we

---

analyze how nearest-neighbor attractive Coulomb interactions and flat-band structures influence superconductivity from the standpoint of magnetic properties. The observed enhancement of the spin structure factor near the *Γ* point in the Brillouin zone indicates that enhanced antiferromagnetic correlations are essential for enhancing the superconducting transition temperature and for stabilizing chiral NN-d+id wave. Through these investigations, our numerical findings not only contribute to a more comprehensive understanding of strongly correlated systems such as twisted bilayer graphene, but also provide guidance for identifying twist-angle systems with potentially higher superconducting transition temperatures.

# 1. Introduction

In 2018, the discovery of correlated insulator and superconductivity in the torsion-angle bilayer graphene system marked a new breakthrough in the study of the physical mechanism of unconventional superconductors, which further stimulated people's interest in unconventional superconductors and opened the door to the study of torsion-angle systems[1,2]. Similar to Cu-based and Fe-based superconductors, it is also a typical strongly correlated system, which shows the characteristics of Mott correlated insulator when undoped, and two obvious arch-shaped superconducting regions appear on both sides of the half-filled Mott correlated insulator when doped, and the conductance of these regions shows a V-shaped trend of[3] and a small coherence length[1,4]. In addition, there are strange metal[5,6] and quantum anomalous Hall effect[7]. Twisted bilayer graphene is more unique in that it forms Moire patterns with different primitive cell sizes in real space as the twist angle changes. The Fermi velocity of Dirac fermions varies with the twist angle. The electronic States near each valley of each graphene layer hybridize with the corresponding electronic States of the other layer. When the twist angle is close to 1.1 °, the electronic structure shows a flat band property. With the increase of carrier concentration, the chemical potential of the system is located near the flat band, and the correlation effect between electrons is significantly enhanced[8–10]. Twisted bilayer graphene is considered to be an ideal system for studying the microscopic mechanism of unconventional superconductivity because of its advantages of no disorder caused by doping and controllable carrier concentration.

For these novel correlated States of the twisted bilayer graphene system, Lu et al.[4] used the four-terminal method to measure the bilayer graphene system with a twist angle close to 1.1 ° and obtained a more detailed phase diagram. The results show that there are integer electrons or holes in the Mohr primitive cell, that is, the $(v = \pm1, \pm2, \pm3)$ resistance has an obvious peak when all non-zero integer Mohr bands are filled. By carefully analyzing the temperature dependence of the conductance, it is found that the system is in the correlated insulator state at this time due to a series of peaks in the conductance. In addition, a more detailed physical phase diagram can be drawn according to the measurement of the resistance value. In addition to the correlated insulator state, the most prominent region in the phase diagram is the critical superconducting state region of the correlated state. For different doping regions and different temperatures, these correlated States will also show different States. In order to have a more comprehensive understanding of the superconducting properties and correlated States of the twisted bilayer graphene system, Cao et al. Successfully prepared various twisted bilayer graphene systems[11]. The experimental results show that, in addition to the superconductivity of the twisted bilayer graphene system with a twist angle of 1.08 °, superconductivity also exists in other bilayer graphene systems with a small twist angle. The relationship between the superconducting critical transition temperature and the twist angle under the optimal doping is observed. It is found that the superconducting critical transition temperature of the bilayer graphene with a twist angle close to 1.08 ° is higher, which indicates that different twist angles have a greater impact on the superconducting critical transformation temperature.

The first problem encountered in the theoretical study of superconductivity in the torsion bilayer graphene system is the sharp increase of the number of atoms in the primitive cell caused by the symmetry of the torsion bilayer graphene. Different from the highly ordered single-layer graphene structure (single-layer graphene satisfies $C_6$ symmetry), the interlayer coupling of twisted double-layer graphene changes with the change of twisting angle, which leads to the continuous change of interlayer symmetry. The most remarkable feature is that when the two layers of graphene are twisted, the lattices of the upper layer of graphene and the lower layer of graphene no longer coincide, and the primitive cell gradually expands as the twist angle becomes smaller and smaller. Taking the twisted bilayer graphene lattice with a twisted angle of $\theta = 1.08°$ as an example, there are as many as 11164 atoms in the primitive cell[12]. Considering the reliability of the theoretical results, the numerical simulation should be carried out in a supercell containing multiple primitive cells, but any theoretical method can not simulate the corresponding real model of the system. Therefore, the key to solve this problem is to construct an effective theoretical model based on a clear understanding of the band structure of twisted bilayer graphene. For the band structure of twisted bilayer graphene, it is noted that a very small twist angle will cause lattice dislocations and Moire fringes with local periods, and it can be predicted that two layers of atoms will form a very narrow flat band opposite to each other in the Moire fringes[13]. This flat band feature is also confirmed by the results of first-principles studies. Lucignano et al. Constructed a twisted bilayer graphene structure with a twist angle of 1.08 °[14], and simulated the band structure of the system by density functional theory. The optimized band structure of the twisted bilayer graphene forms a very narrow plane near the Fermi surface $E$ = 0.0 meV, and the corresponding band width is about 12 meV, which is consistent with the experimental band width. Based on the Wannier orbital symmetry analysis of twisted bilayer graphene, yuan and Fu[15,16] have recently demonstrated that the effective two-orbital Hubbard model in honeycomb lattice can reflect the basic physical properties of twisted bilayer graphene, such as electronic structure and Coulomb interaction effect. This model provides a useful theoretical basis for the study of the reconstruction of Dirac fermions by the torsion angle and the charge carrier concentration in torsion bilayer graphene.

Secondly, the electron pairing form of twisted bilayer graphene superconductors is also controversial. Previous studies have predicted that the half-filled state is a Mott insulator, while the electron-doped state may be a topological superconducting state, considering both the flat band structure and the repulsive electron interaction[17]. More detailed theoretical studies predict that the electron pairing of twisted bilayer graphene may be chiral $d + id$ wave[18–22], extended s-wave and p-wave, and f-wave[23–29], and more consensus is that it is most likely to be chiral $d + id$ wave pairing[30]. In addition to these important previous results, more and more research works have shown that the electron-phonon interaction plays an important role in the superconductivity of twisted bilayer graphene systems[31–38]. For example, Chen et al.[31] used angle-resolved photoemission spectroscopy with micrometer resolution to observe the flat band structure in twisted bilayer graphene in the superconducting state. These flat band structures have a uniform band spacing of about 150 meV, while the system does not have this behavior in the non-superconducting state, indicating that the system has strong electron-phonon coupling. It is also confirmed that the formation of the flat band structure of the twisted bilayer graphene in the superconducting state originates from the strong coupling between the transverse optical phonon mode at the $K$ point of graphene and electrons.

Motivated by the above series of research works, in this paper, we use the large-scale unbiased constrained path quantum Monte Carlo method to study the intrinsic correlation between magnetism and superconductivity in the torsion bilayer graphene system of the effective two-orbital Hubbard model, and discuss the control effect of the attractive nearest neighbor Coulomb interaction on superconductivity and the influence of magnetism on superconductivity from the perspective of electron-phonon coupling. The results show that the electron-phonon coupling can strongly enhance the superconductivity of the system, and from a deeper point of view, the enhancement of antiferromagnetic order is a prerequisite for the control of superconductivity. On the other hand, by constructing the effective two-orbital Hubbard model of bilayer graphene with different torsion angles, the control effect of torsion angle on superconductivity and the correlation between magnetism and superconductivity are discussed. The results show that a smaller torsion angle is beneficial to enhance the superconductivity of the system, and the coordination effect between antiferromagnetic order and superconductivity is confirmed again.

## 2. Model and calculation method

### 2.1 Theoretical model

This work uses the effective two-orbital Hubbard model proposed by [15,16,39] to describe the structure of twisted bilayer graphene. The model includes the kinetic energy part $\hat{H}_t$ and the interaction part $\hat{H}_{int}$. The Hamiltonian $\hat{H}_t$ includes the intra-orbital $\hat{H}_0$, the symmetry breaking terms $\hat{H}_1$ and $\hat{H}_2$, in which the two symmetry breaking terms are introduced to break the $SU(4)$ and $U(1)$ symmetries, respectively. The Hamiltonian $\hat{H}_t$ is described by

$$\hat{H}_t = \hat{H}_0 + \hat{H}_1 + \hat{H}_2 \hat{H}_0$$
$$\hat{H}_0 = \sum_{\langle i,j\rangle,\sigma} t_1[\hat{c}_{i,\sigma}^\dagger \cdot \hat{c}_{j,\sigma} + h.c.] + \sum_{\langle i,j\rangle',\sigma} t_2[\hat{c}_{i,\sigma}^\dagger \cdot \hat{c}_{j,\sigma} + h.c.]$$
$$\hat{H}_1 = \sum_{\langle i,j\rangle',\sigma} t_{2'}[(\hat{c}_{i,\sigma}^\dagger \times \hat{c}_{j,\sigma})_z + h.c.] \quad (2,1)$$
$$= -i\sum_{\langle i,j\rangle',\sigma} t_{2'}(\hat{c}_{i_+,\sigma}^\dagger \hat{c}_{j_+,\sigma} - \hat{c}_{i_-,\sigma}^\dagger \hat{c}_{j_-,\sigma}) + h.c.$$
$$\hat{H}_2 = \sum_{\langle i,j\rangle,\sigma} t_1'[\hat{c}_{i,\sigma}^\dagger \cdot e_{ij}^\| e_{ij}^\| \cdot \hat{c}_{j,\sigma} - \hat{c}_{i,\sigma}^\dagger \cdot e_{ij}^\perp e_{ij}^\perp \cdot \hat{c}_{j,\sigma} + h.c.],$$

Where $\hat{c}_{i,\sigma} = (\hat{c}_{i,x,\sigma}, \hat{c}_{i,y,\sigma})^T$, the operator $\hat{c}_{i,x(y),\sigma}(\hat{c}_{i,x(y),\sigma}^\dagger)$ represents the annihilation (creation) of a $p_{x(y)}$ electron with spin $\sigma(\sigma=\uparrow,\downarrow)$ located at the $i$ site. As shown in the Fig. 1(a), $t_1$ and $t_2$ represent the electron hopping integral strength of the nearest neighbor and the fifth neighbor within the orbital, respectively. The specific form of $\hat{c}_{\pm,\sigma} = (\hat{c}_{x,\sigma} \pm i\hat{c}_{y,\sigma})/\sqrt{2}$ is $p_{x,\sigma} \pm ip_{y,\sigma}$, and obeys the fundamental rule of chirality. In the formula, $\hat{H}_1$ breaks the $SU(4)$ symmetry and thus decomposes into $U(1) \times SU(2)$ symmetry. Here $U(1)$ symmetry refers to conservation of orbital chirality and $SU(2)$ symmetry refers to spin rotation symmetry. The $e_{ij}^{\parallel,\perp}$ in the $\hat{H}_2$ term of the Hamiltonian represents the parallel and perpendicular unit vectors connecting with the nearest neighbors. The specific kinetic energy expressions of the $\hat{H}_2$ term of the Hamiltonian along the three directions of the nearest-neighbor site connection are

$$\hat{c}_{ix}^\dagger \hat{c}_{jx} - \hat{c}_{iy}^\dagger \hat{c}_{jy} - \frac{1}{2}(\hat{c}_{ix}^\dagger \hat{c}_{jx} - \hat{c}_{iy}^\dagger \hat{c}_{jy}) - \frac{\sqrt{3}}{2}(\hat{c}_{ix}^\dagger \hat{c}_{jy} + \hat{c}_{iy}^\dagger \hat{c}_{jx}) - \frac{1}{2}(\hat{c}_{ix}^\dagger \hat{c}_{jx} - \hat{c}_{iy}^\dagger \hat{c}_{jy}) + \frac{\sqrt{3}}{2}(\hat{c}_{ix}^\dagger \hat{c}_{jy} + \hat{c}_{iy}^\dagger \hat{c}_{jx})$$

Here $\hat{H}_2$ further breaks the $U(1)$ symmetry. It should be pointed out that the jump parameters in this paper are different from those in references [15,16]. In this paper, the hopping amplitude $t_1$ is taken as the unit energy, and the other hopping parameters are set to $t_2 = 0.025$ and $t_1' = t_2' = 0.1$.

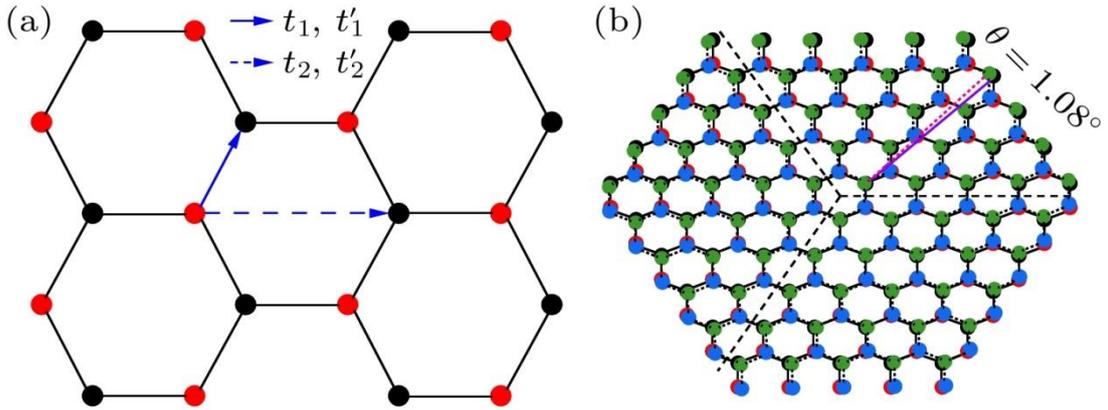

**Figure 1.** (a) Schematic diagram illustrating the electron hopping terms of effective two-orbital Hubbard model on the twisted bilayer graphene. The black (red) dots represent sublattice A (B), with each lattice point containing two orbitals. The hopping integrals $t_1$ and $t_1'$ correspond to nearest-neighbor interactions, while $t_2$ and $t_2'$ represent the fifth-nearest-neighbor interactions. (b) Schematic of the lattice structure with a twist angle of 1.08° and lattice size $L = 5$. Black and red points correspond to the inequivalent carbon atoms A and B in the first layer, while green and blue points represent the inequivalent atoms $A_1$ and $B_1$ in the second layer. The three black curves denote the periodic boundary lengths of the structure.

The interaction part includes the on-site Coulomb repulsive interaction term $\hat{H}_U$ and the intra-orbital nearest-neighbor Coulomb interaction term $\hat{H}_V$, and the Hamiltonian $\hat{H}_{\text{int}}$ of this part is described by

$$\hat{H}_{int} = \hat{H}_U + \hat{H}_V = U\sum_{i,\beta}\hat{n}_{i\beta\uparrow}\hat{n}_{i\beta\downarrow} + \sum_{<i,j>,\sigma} V\hat{n}_{i,\sigma}\hat{n}_{j,\sigma} \qquad (2,2)$$

Where $\beta$ denotes the $p_{x(y)}$ orbital, $<i,j>$ denotes the nearest-neighbor site, $c$ denotes the on-site Coulomb repulsive interaction strength, and $V$ denotes the intra-orbital nearest-neighbor Coulomb interaction strength.

2.2 Calculation method and physical quantity

Before discussing the calculation method and results, it should be noted that the numerical simulations in this paper are carried out in a periodic hexagonal lattice. As shown in Fig. 1(b), the figure represents a twisted bilayer graphene model with a side length of $L$ = 5 and a twist angle of 1.08 °, which contains $4\times 3L^2$ atoms. In order to solve this multi-atom Hubbard model, the large-scale unbiased constrained path quantum Monte Carlo method, which can accurately solve the ground state information of correlated electron systems, is used. The basic strategy of this method is to obtain the ground state wave function $|\Psi_g\rangle$ of the correlated electron system by projecting an initial wave function $|\Psi_T\rangle$ through a multi-branch random walk in Slater determinant space, and the projection equation can be described as

$$|\Psi_g\rangle = \lim_{\tau\to\infty} e^{-\tau\hat{H}}|\Psi_T\rangle, \qquad (2,3)$$

Where $\hat{H}$ denotes the Hamiltonian of the system, and $\tau$ is real and greater than the $0$. Numerically, the projection equation can be implemented in an iterative manner, i.e.

$$|\Psi^{(n+1)}\rangle = e^{-\Delta\tau\hat{H}}|\Psi^{(n)}\rangle \qquad (2,4)$$

At the initial iteration $|\Psi^{(0)}\rangle = |\Psi_T\rangle$, for very small $\Delta\tau$, the Trotter-Suzuki $(TS)$ decomposition of the[40,41] can be used, and the (4) can be rewritten as

$$|\Psi^{(n+1)}\rangle \approx e^{-\Delta\tau\hat{H}_t/2} e^{-\Delta\tau\hat{H}_{int}} e^{-\Delta\tau\hat{H}_t/2}|\Psi^{(n)}\rangle \qquad (2,5)$$

Where $\hat{H}_t$ is the quadratic Fermi operator, which acts on the initial wave function in the form of an exponential, simply transforming the Slater determinant corresponding to the wave function into another Slater determinant. For the $\hat{H}_{int}$ Fermi operator with quartic operator form, we need to transform the operator $e^{-\Delta\tau\hat{H}_{int}}$ into quadratic operator form by using the discrete Hubbard-Stratonovich (HS) transformation, which can be found in reference [42].

The final iteration equation (4) through TS decomposition and HS transformation is expressed as

$$\begin{aligned}|\Psi^{(n+1)}\rangle &= e^{-\Delta\tau\hat{H}}|\Psi^{(n)}\rangle \\ &= e^{-\Delta\tau\hat{H}_t/2}e^{-\Delta\tau\hat{H}_{int}}e^{-\Delta\tau\hat{H}_t/2}|\Psi^{(n)}\rangle \\ &= e^{-\Delta\tau\hat{H}_t/2}\sum_{x_i,x_i^n=\pm 1}\prod_i \rho(x_i)e^{-\Delta\tau U(\hat{n}_{i\uparrow}+\hat{n}_{i\downarrow})/2}e^{\lambda x_i(\hat{n}_{i\uparrow}-\hat{n}_{i\downarrow})} \\ &\cdot \prod_{ij}\rho(x_{ij}^1)\rho(x_{ij}^2)\rho(x_{ij}^3)\rho(x_{ij}^4)e^{\Delta\tau V(\hat{n}_{i\sigma}+\hat{n}_{j\sigma})/2}e^{\lambda' x_i(\hat{n}_{i\sigma}-\hat{n}_{j\sigma})}e^{-\Delta\tau\hat{H}_t/2}|\Psi^{(n)}\rangle\end{aligned} \quad (2,6)$$

At this point, the projection operator $e^{-\Delta\tau\hat{H}}$ becomes a complex quadratic operator form, although the operator is complex, but the advantage is to overcome the difficulty of solving four operators. After the above treatment, the equation (6) can be regarded as the summation of multiple auxiliary fields, and the corresponding expression is

$$|\Psi^{(n+1)}\rangle = P(x)\hat{B}(x)|\Psi^{(n)}\rangle \quad (2,7)$$

In formula $P(x) = \sum_{x_i,x_i^n=\pm 1}\prod_i \rho(x_i)\prod_{ij}\rho(x_{ij}^1)\rho(x_{ij}^2)\rho(x_{ij}^3)\rho(x_{ij}^4)$ Stands for the probability density function, $\hat{B}^\sigma(x) = \hat{B}_K^\sigma \hat{B}_I^\sigma$ ($\sigma = \uparrow, \downarrow$) stands for the auxiliary field propagator of the HS transformation, $\hat{B}_K^\sigma$ is the propagator of the kinetic energy part, and $\hat{B}_I^\sigma(x)$ is the propagator of the potential energy part. The Monte Carlo method is a multidimensional summation of the auxiliary field $x_i$. When the wave function converges, the expectation value of the $\mathcal{O}$ of the physical quantity in the corresponding system is usually obtained based on the backpropagation[43] technique, that is,

$$\langle\mathcal{O}\rangle_{BP} = \frac{\langle\Psi_T|e^{-l_{BP}\Delta\tau\hat{H}}\mathcal{O}|\Psi_0\rangle}{\langle\Psi_T|e^{-l_{BP}\Delta\tau\hat{H}}|\Psi_0\rangle} \quad (2,8)$$

Where $l_{BP}$ is the number of return steps, which is set in the numerical simulation process $l_{BP} = 40$.

From the above introduction, it can be seen that the choice of the initial wave function will be the key factor affecting the accuracy of the numerical simulation results. The advantage of the constrained path quantum Monte Carlo method is that the statistical error obtained is consistent with the results whether the free electron wave function is used as the initial wave function or other types of wave functions (such as the unrestricted Hartree-Fock wave function)[44–46]. In terms of simulation technology, the Monte Carlo method is to generate a sampling sample in the space of Slater determinants by random multi-branch walk, that is, the wave function is written as $|\Psi^{(n)}\rangle = \sum_k \chi_k |\phi_k^n\rangle$. At the same time, considering that the ground state wave function of fermions needs to satisfy the antisymmetry, for the set $\{|\phi\rangle\}$ of Slater determinants representing the ground state wave function, the antisymmetry means that there is another set $\{-|\phi\rangle\}$ of Slater determinants that can also correctly represent the ground state wavefunction. With the increase of sampling times, the Monte Carlo method will inevitably have a sign problem in numerical simulation. The core idea and advantage of the constrained path quantum Monte Carlo method is to limit the sample to a sample space by defining a reasonable interface N'. In numerical calculation, the method requires that the sample $|\phi_k^{(n)}\rangle$ and the trial wave function $|\Psi_T\rangle$ have a positive overlap integral $\langle \Psi_T | \phi_k^{(n)} \rangle > 0$, which is the so-called constrained path approximation. It is this approximation that ensures the fast convergence and accuracy of the results.

In order to fully understand the superconducting properties in twisted bilayer graphene, the intra-orbital electron pairing correlation function is calculated for various pairing forms, and the pairing correlation function is defined as follows:

$$P_\alpha(R = |\mathbf{R}|) = \frac{1}{3L^2} \sum_{\mathbf{i}} \langle \Delta_\alpha^\dagger(\mathbf{i}+\mathbf{R}) \Delta_\alpha(\mathbf{i}) \rangle \tag{2,9}$$

Where $\Delta_\alpha^\dagger(\mathbf{i})(\Delta_\alpha(\mathbf{i}))$ is the electron pairing creation (annihilation) operator with pairing symmetry $\alpha$. The creation operators for the singlet and triplet States can be written as

$$\Delta_\alpha^\dagger(\mathbf{i}) = \frac{1}{\sqrt{N_\alpha}} \sum_{l\beta} f_\alpha(\boldsymbol{\delta}_l)(\hat{c}_{\mathbf{i},\beta,\uparrow}^\dagger \hat{c}_{\mathbf{i}+\boldsymbol{\delta}_l,\beta,\downarrow}^\dagger \pm \hat{c}_{\mathbf{i},\beta,\downarrow}^\dagger \hat{c}_{\mathbf{i}+\boldsymbol{\delta}_l,\beta,\uparrow}^\dagger), \tag{2,10}$$

Where $f_\alpha(\boldsymbol{\delta}_l)$ is the shape factor used to distinguish between different types of pairing. −(+) corresponds to singlet (triplet) pairing. Depending on the type of pairing, the vector $\boldsymbol{\delta}_l$ represents the connection of nearest neighbors or next nearest neighbors of the same kind of lattice. $N_\alpha$ is the normalization coefficient, for nearest-neighbor pairing $N_\alpha = 3$, next-nearest-neighbor pairing $N_\alpha = 6$. All possible nearest-neighbor and next-nearest-neighbor pairings are considered. These pairings are NN-s symmetric wave, NN-d + id symmetric wave, NN-p + ip symmetric wave, NNN-d + id symmetrical wave, NNN-p + ip symmetrical wave and NNN-f symmetrical wave[47,48]. The form factor $f_\alpha(\boldsymbol{\delta}_l)$ of these pairing symmetries is shown in Fig. 2, and its specific definition is

$$f_{NN,s}(\boldsymbol{\delta}_l) = 1, \quad f_{NN,d+id}(\boldsymbol{\delta}_l) = e^{i(l-1)\frac{2\pi}{3}},$$

$$f_{NN,p+ip}(\boldsymbol{\delta}_l) = e^{i(l-1)\frac{2\pi}{3}}, f_{NNN,d+id}(\delta_{l'}) = e^{i(l'-1)\frac{2\pi}{3}}, \quad (2,11)$$

$$f_{NNN,p+ip}(\delta_{l'}) = e^{i(l'-1)\frac{\pi}{3}}, f_{NNN,f}(\delta_{l'}) = e^{i\cdot\frac{1+(-1)^{l'}}{2}\pi}$$

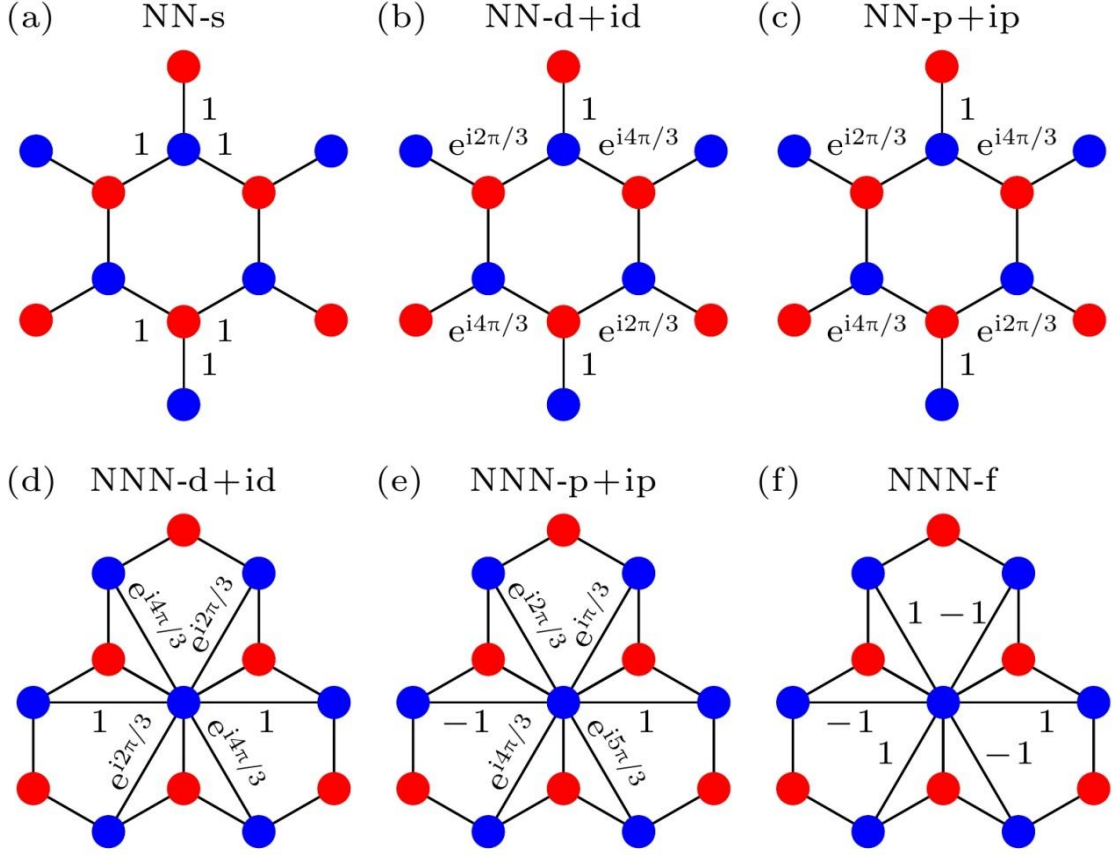

**Figure 2.** Schematic diagrams of various intra-orbital electron pairing symmetry: (a) NN-s-wave symmetry; (b) NN-d + id-wave symmetry; (c) NN-p + ip-wave symmetry; (d) NNN-d + id-wave symmetry; (e) NNN-p + ip-wave symmetry; (f) NNN-f-wave symmetry.

In the equation (11), the vector $\boldsymbol{\delta}_l (l = 1, 2, 3)$ represents the direction of the nearest neighbor lattice, and $\delta_{l'} (l' = 1, 2, 3, 4, 5, 6\$)$ represents the direction of the next nearest neighbor lattice point.

Considering that the electron pairing correlation function may be easily affected strongly by the non-interacting part in the Hamiltonian,[48–50], the effective pairing correlation function is also calculated in this paper, which is defined as

$$V_\alpha(R) = P_\alpha(R) - \bar{P}_\alpha(R) \qquad (2,12)$$

Where $\tilde{P}_\alpha(R)$ is the uncorrelated single-particle contribution, which can be obtained by decoupling $\langle \hat{c}^\dagger_{i\downarrow} \hat{c}_{j\downarrow} \hat{c}^\dagger_{k\uparrow} \hat{c}_{l\uparrow} \rangle$ into $\langle \hat{c}^\dagger_{i\downarrow} \hat{c}_{j\downarrow} \rangle \langle \hat{c}^\dagger_{k\uparrow} \hat{c}_{l\uparrow} \rangle$. The system-dominated pairing form can be determined by the enhancement (suppression) trend of this effective pairing correlation function. In order to be able to reflect the effect of the interaction part on the long-range order of the electron pairing symmetry, the long-range average effective pairing correlation function can be expressed as

$$\overline{V_\alpha}(R \geq 3) = \frac{1}{N_1} \sum_{R \geq 3} V_\alpha(R) \tag{2,13}$$

Where $R$ represents the distance between sites, and $N_1$ represents the number of site pairings of the electron pairing distance $R \geq 3$.

In the study of the physical properties of various superconductors, it is generally believed that magnetism is closely related to superconductivity, so the magnetic order in twisted bilayer graphene is also calculated in this paper, and the spin structure factor $S(q)$ in the wave vector space is defined as

$$S(q) = \frac{1}{12L^2} \sum_{d,d'=x,y} \sum_{i,j} \epsilon d, d' e^{iq(i_d - j_{d'})} \langle \hat{S}_{i,d} \hat{S}_j, d' \rangle \tag{2,14}$$

Where $\hat{S}_{i,d} = \hat{n}_{i,d,\uparrow} - \hat{n}_{i,d,\downarrow}$; in case of $d = d', \epsilon = 1$; in case of $d \neq d', \epsilon = -1$, Finally, we use the constrained path quantum Monte Carlo method to simulate the torsion bilayer graphene model. The closed shell filling of electrons and the free electron wave function are selected as the initial wave function. The main parameters in the simulation are: the number of random walks 9000, the projection step $\Delta = 0.04$. Before measuring the physical quantity, 1280 steps of Monte Carlo sampling are carried out. When measuring the physical quantity, 320 steps of Monte Carlo sampling are set. 40 independent tests ensure the statistical independence. When the simulation reaches equilibrium, the expected value of the physical quantity can be solved based on the feedback technique.

## 3. Calculation results and discussion

3.1 Influence of the nearest neighbor Coulomb interaction on the superconducting state

This subsection focuses on the influence of the nearest neighbor Coulomb interaction strength on the superconducting pairing symmetry of the twisted bilayer graphene system, and Fig. 3 shows the evolution curves of the pairing correlation function of various electron pairing symmetries with the lattice size of $L = 5$ on the pairing distance $R \geqslant 3$ for different Coulomb interaction strengths. Fig. 3(a) and Fig. 3(b) correspond to the cases of $U = 2.0, V = 0.0$ and $U = 0.0, V = -0.3$, respectively, at electron filling concentration $\langle n \rangle = 0.933$, while Fig. 3(c) and Fig. 3(d) correspond to the cases of $U = 2.0, V = 0.0$ and $U = 0.0, V = -0.3$, respectively, at electron filling concentration $\langle n \rangle = 0.893$. It can be clearly observed from the figure that for the case of only on-site Coulomb interaction strength $U = 2.0$ (Fig. 3(a) and Fig. 3(c)) or nearest neighbor attractive Coulomb interaction strength $V = -0.3$ (Fig. 3(b) and Fig. 3(d)), the strength of triplet NNN-f wave pairing symmetry is always at the highest position over the whole long range distance $R \geqslant 3.0$. Based on the current numerical simulation results, it seems that the pairing symmetry of electrons in the twisted bilayer graphene system is triplet NNN-f wave pairing. However, an important problem is that previous quantum Monte Carlo studies of superconductivity in graphene systems have pointed out that the pairing symmetry of electrons is strongly affected by the non-interacting part of the Hamiltonian[48–50], so the conclusions drawn from the pairing correlation function are not very credible.

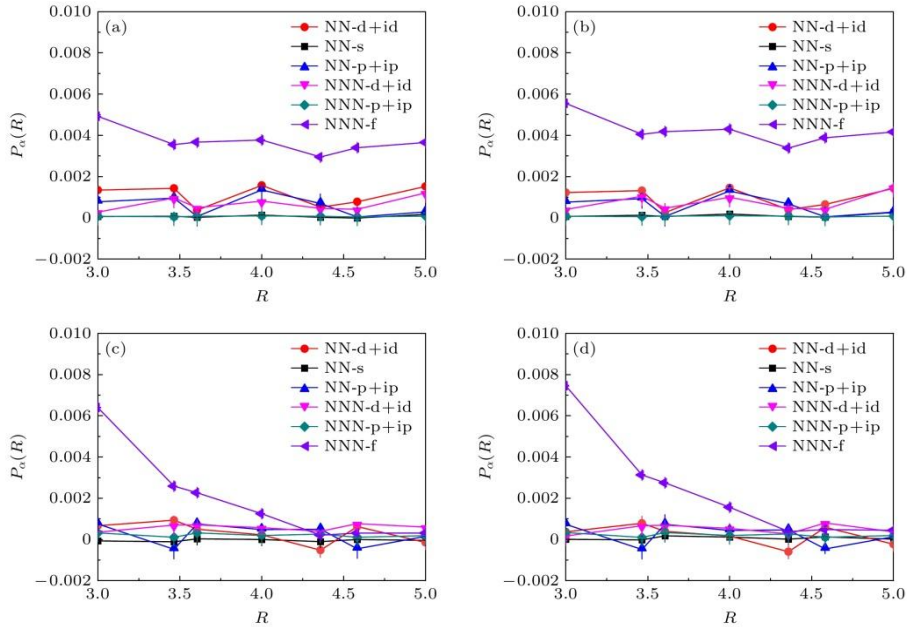

**Figure 3.** Pairing correlation functions $P_\alpha(R)$ as a function of pairing distance $R$ for various electron pairing symmetries in a lattice of size $L = 5$: **(a)** $U = 2.0$, $V = 0.0$, $\langle n \rangle = 0.933$; **(b)** $U = 0.0$, $V = -0.3$, $\langle n \rangle = 0.933$; **(c)** $U = 2.0$, $V = 0.0$, $\langle n \rangle = 0.893$; **(d)** $U = 0.0$, $V = -0.3$, $\langle n \rangle = 0.893$.

In order to further confirm whether the non-interacting part of the Hamiltonian has a strong influence on the electron pairing symmetry, Fig. 4(a) and Fig. 4(b) show the evolution curves of various electron pairing symmetries with the pairing distance $R$ in the absence of interaction. The numerical simulation results in the figure show that the pairing strength of the triplet NNN-f wave is still greater than that of other types of electron pairing symmetry in the case of electron filling concentration $\langle n \rangle = 0.933$ and $\langle n \rangle = 0.893$. By comparing the results of Fig. 3 and Fig. 4, it can be found that the correlation functions of various pairing forms have the same trend with the long-range distance *R*, so it is impossible to analyze the influence of interaction part on the symmetry of electron pairing. Based on this result, it is concluded from the correlation function that the pairing form dominated by triplet NNN-f wave pairing symmetry should be caused by the electronic structure of the non-interacting part. Therefore, the non-interacting part conceals the effective information of the interacting part on the electron pairing symmetry, that is, the electron pairing symmetry of the twisted bilayer graphene cannot be accurately analyzed from the correlation function.

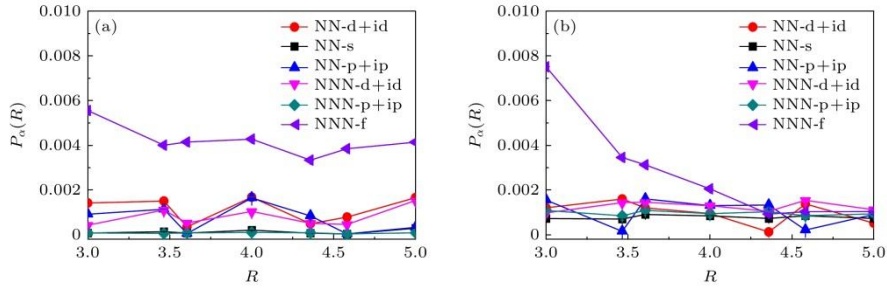

**Figure 4. Pairing correlation functions $P_\alpha(R)$ as a function of pairing distance *R* for various electron pairing symmetries in a non-interacting system with lattice size $L = 5$: (a) $\langle n \rangle = 0.933$; (b) $\langle n \rangle = 0.893$.**

Based on the previous discussion, considering the influence of the non-interacting part on the pairing symmetry of electrons, the next step is to analyze the pairing symmetry of electrons based on the effective pairing correlation function. Recent angle-resolved photoemission spectroscopy has revealed the variation of hole and spin energy bands with doping concentration in one-dimensional copper-based superconductors, and found an obvious hole folding branch, which was pointed out in reference [51] to originate from the nearest neighbor attractive Coulomb interaction caused by electron-phonon coupling. In addition, the theoretical results obtained by using the variational non-Gaussian exact diagonalization method in simulating the one-dimensional Hubbard model are quantitatively compared with the above experimental results, and the attractive interaction strength between neighboring electrons is obtained to be in good agreement with the experimental observations in the reasonable electron-phonon coupling range[52]. Secondly, the numerical results of the checkerboard tetragonal lattice using the quantum Monte Carlo method reveal that the d-wave superconducting state is strongly enhanced by the nearest neighbor attractive Coulomb interaction[53]. Therefore, the specific manifestation of electron-phonon coupling can be regarded as the attractive interaction between neighboring electrons, which may play an important role in understanding the mechanism of unconventional superconductors. Inspired by this series of works, this paper systematically studies the regulation of electron-phonon coupling on the superconductivity of twisted bilayer graphene system by the attractive Coulomb interaction between neighboring electrons. The functional dependence of the average effective pairing correlation function $\overline{V}_\alpha(R \geqslant 3)$ on the strength of the nearest neighbor attractive Coulomb interaction $V$ is shown for various pairing forms Fig. 5. The numerical results of Fig. 5(a) and Fig. 5(b) are also obtained by simulating the lattice size of $L = 5$, and the electron filling concentrations are also $\langle n \rangle = 0.933$和$\langle n \rangle = 0.893$. First of all, in the case of nearest neighbor Coulomb interaction strength $V = 0.0$, it can be easily observed that the average effective pairing correlation strength corresponding to the chiral $NN\text{-}d + id$ wave is significantly larger than that of other electron pairing forms, which is obviously completely different from the triplet NNN-f wave dominated pairing form obtained by Fig. 3 based on the pairing correlation function, which not only confirms that the non-interacting part does have a strong impact on the electron pairing symmetry, but also indicates that the twisted angle bilayer graphene system. Secondly, the numerical simulation results of Fig. 5 also show that the average effective pairing correlation function of all kinds of electron pairing forms increases with the increase of the strength of the nearest neighbor attractive Coulomb interaction. More importantly, the chiral $NN\text{-}d + id$ wave electron pairing symmetry is always dominant and has the largest increase. It should be pointed out that, compared with the average effective pairing correlation strength of the chiral $NN\text{-}d + id$ wave pairing symmetry when the nearest neighbor Coulomb interaction strength is $V = 0.0$, when the attractive Coulomb interaction is present and the strength is increased to $V = -0.3$, For the case where the electron filling concentration ⟨n⟩=0.933, the pairing strength of the chiral $NN\text{-}d + id$ wave increases by a factor of 2.106; In the case of electron filling concentration $\langle n \rangle = 0.893$, the pairing strength of the chiral $NN\text{-}d + id$ wave increases by a factor of 1.514, which fully indicates that the strength of the nearest neighbor attractive Coulomb interaction can drastically enhance the chiral $NN\text{-}d + id$ superconducting state.

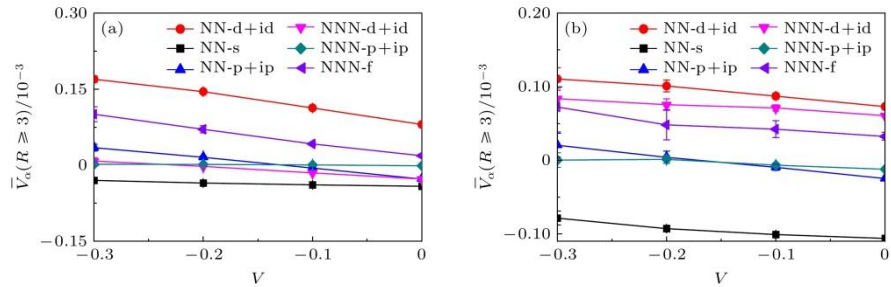

**Figure 5.** Average effective pairing correlation functions $\overline{V}_\alpha(R \geqslant 3)$ as a function of nearest-neighbor Coulomb interaction $V$ for various electron pairing symmetries in a system with on-site Coulomb interaction strength $U = 2.0$ and lattice size $L = 5$: (a) $\langle n \rangle = 0.933$; (b) $\langle n \rangle = 0.893$.

Considering the lattice size effect, the numerical simulation is carried out on a larger lattice scale. Taking the lattice size $L = 6$ as an example, Fig. 6(a) and Fig. 6(b) show the effective pairing correlation function of chiral NN-d+id wave pairing symmetry as a function of the long-range distance $R \geqslant 3$ for the case of on-site Coulomb interaction strength $U = 2.0$, where the electron filling concentration $\langle n \rangle = 0.954$ (Fig. 6(a)) and the electron filling concentration $\langle n \rangle = 0.926$ (Fig. 6(b)) are selected. Based on the data in this figure, it can be observed that the effective pairing correlation function of the chiral NN-d+id wave pairing symmetry increases with the increase of the nearest neighbor attractive Coulomb interaction strength $V$ over the whole long range, especially near the half-filled electron concentration $\langle n \rangle = 0.954$, where the change of the chiral NN-d+id wave is the most significant. The consistency of the results of Fig. 5 and Fig. 6 indicates that the chiral NN-d+id superconducting state is effectively enhanced by the nearest neighbor attractive Coulomb interaction strength $V$.

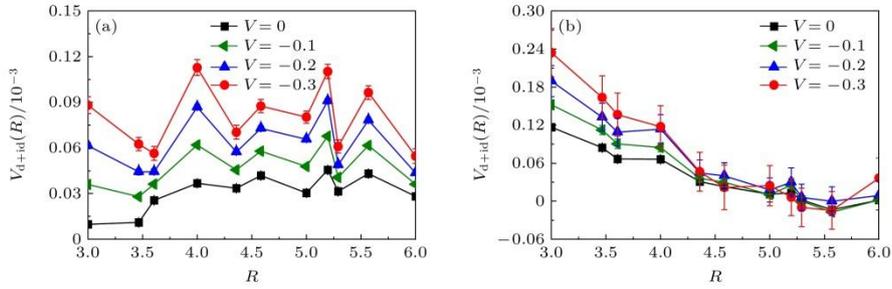

Figure 6. **Effective pairing correlation functions $V_{d+id}$ as a function of long-range pairing distance $R$ for the chiral NN-$d + id$-wave pairing symmetry in a system with on-site Coulomb interaction strength $U = 2.0$ and lattice size $L = 6$: (a) $\langle n \rangle = 0.954$; (b) $\langle n \rangle = 0.926$.**

Magnetism and superconductivity are widely considered to be closely related in unconventional superconducting States. Antiferromagnetic fluctuations play an important role in modulating superconductivity, such as in the typical high temperature cuprate superconductor[54–57] and the iron-based superconductor[58–61]. Based on the band structure of twisted bilayer graphene, it is found that the electrons in each orbital valley hybridize with the electrons in the other orbital and show Mott insulator characteristics when they are not filled, which indicates that the system is similar to copper-based superconductors and iron-based superconductors, and is also a typical strongly correlated electron system, so it can be predicted that magnetism also has an important impact on superconductivity in this system. Considering the correlation of magnetism and superconductivity, the Fig. 7 shows the evolution of the spin structure factor $S(q)$ along the high-symmetry line of the first Brillouin zone. The data simulation of this figure is carried out in the on-site Coulomb interaction strength $U = 2.0$, the lattice size $L = 5$, and the electron filling concentration $\langle n \rangle = 0.933$ (Fig. 7(a)), $\langle n \rangle = 0.893$ (Fig. 7(b)), respectively. It can be clearly observed from Fig. 7 that the spin structure factor $S(q)$ shows an obvious increasing trend at $\Gamma$ point with the increase of the neighbor attractive Coulomb interaction strength $V$, which indicates that the electrons are arranged antiferromagnetically and the system is in a robust antiferromagnetic state. Therefore, this result indicates that the modulation of antiferromagnetic order is an important reason for the enhancement of the superconducting state of chiral NN-d+id.

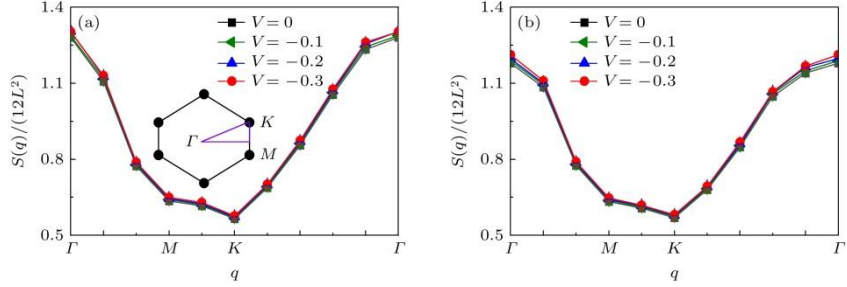

**Figure 7. Spin structure factor $S(q)$ along the high-symmetry lines $\Gamma \to M \to K \to \Gamma$ in the first Brillouin zone for a system with on-site Coulomb interaction strength $U = 2.0$ and lattice size $L = 5$: (a) $\langle n \rangle = 0.933$; (b) $\langle n \rangle = 0.893$. The inset shows the high-symmetry lines in the first Brillouin zone, with the coordinates of *Γ*, *M*, and *K* given by $(0,0)$, $(\frac{2\pi}{3},0)$, and $(\frac{2\pi}{3\sqrt{3}},0)$, respectively. The purple lines in the inset represent the high-symmetry lines.**

3.2 Effect of the flat band structure on the superconducting state

In addition to the discussion of the effect of the nearest neighbor attractive Coulomb interaction on the superconducting state of the twisted bilayer graphene, we also focus on the superconductivity of the twisted bilayer graphene with different angles around 1.08 °. The experimental[8,9,62] and theoretical[14,63–65] results show that the most important feature of twisted bilayer graphene is that the band structure near the Fermi level becomes flatter with the decrease of the twist angle when the twist angle is gradually close to 1. 08 °, which indicates that there is a direct correlation between the twist angle and the band structure. Motivated by this, we can construct different types of flat band structures to study the effect of twist angle on superconductivity by adjusting the electron hopping strength in the breaking term of the effective two-orbital Hamiltonian of twist angle bilayer graphene. Different types of flat band structures can be constructed by synchronously adjusting the strength of $t'_1$ amd $t'_2$ in the equation (1). Shown Fig. 8 are the band structure and density of States of the noninteracting Hamiltonian of twisted bilayer graphene. Fig. 8(a), Fig. 8(c), Fig. 8(e) represent the evolution curve of the band structure along the direction of the high symmetry line of the first Brillouin zone. The method of constructing the band structure here is to fix $t_1 = 1.0$, $t_2 = 0.025$, adjust $t'_1$ and $t'_2$ synchronously. The corresponding electron hopping parameters of Fig. 8(a), Fig. 8(c), Fig. 8(e) are $t'_1 = t'_2 = 0.15$, $t'_1 = t'_2 = 0.10$, $t'_1 = t'_2 = 0.05$. Comparing the band structures in the three subfigures, it can be found that the width of the band structure on the high symmetry line in the $\Gamma \to M$ interval of the first Brillouin zone decreases gradually with the decrease of the electron hopping intensity of $t'_1$ and $t'_2$, that is, the flat-band trend becomes more and more significant. Combined with the results of first-principles calculations and the effective two-orbital Hubbard model[14–16] of twisted bilayer graphene, it is shown that $t'_1 = t'_2 = 0.10$ corresponds to the band characteristics of twisted bilayer graphene with a twist angle of 1. 08 °, $t'_1 = t'_2 = 0.15$ corresponds to the band characteristics of twisted double-layer graphene with a twist angle greater than 1. 08 °, and $t'_1 = t'_2 = 0.05$ corresponds to the band-characteristics of twisted double-layer graphene with a torsion angle less than 1.08°. At the same time, for the band structures with different twist angles, Fig. 8(b), Fig. 8(d), Fig. 8(f) show the functional dependence of the number of States on the energy. The red and blue dashed lines in these subfigures mark the Fermi level positions with electron filling concentrations of $\langle n \rangle = 0.933$ and $\langle n \rangle = 0.893$, respectively.

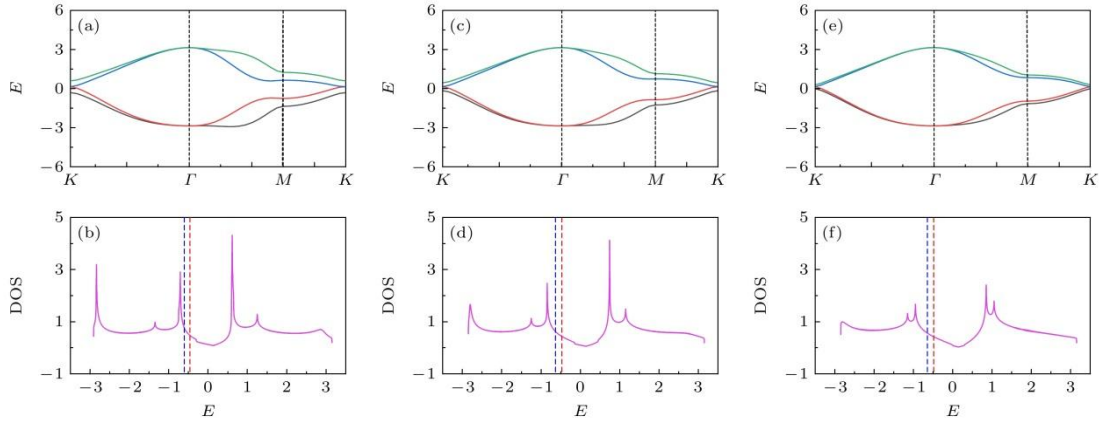

**Figure 8.** Band structure and density of states (DOS) for the non-interacting Hubbard model. Panels (a), (c), and (e) display the band dispersion along the high-symmetry lines in the first Brillouin zone for $t'_1 = t'_2 = 0.15$, $t'_1 = t'_2 = 0.10$, and $t'_1 = t'_2 = 0.05$, respectively. Panels (b), (d), and (f) show the density of states as a function of energy for the same values of $t'_1 = t'_2$. The red and blue dashed lines represent the Fermi level positions corresponding to electron fillings of $\langle n \rangle = 0.933$ and $\langle n \rangle = 0.893$, respectively.

Based on the above discussion, the flat band structures of different types of twisted bilayer graphene are used to discuss the effect of twist angle on the superconducting state. The numerical simulation results of Fig. 9 are the curves of the average effective pairing correlation function $\overline{V}_\alpha(R \geqslant 3)$ as a function of the electron hopping breaking term $t'_{1,2}$ for various types of electron pairing symmetry. The lattice size of the numerical simulation is $L = 5$, the on-site Coulomb interaction strength Is $U = 2.0$ and the nearest neighbor Coulomb interaction strength is $V = 0.0$. The electron filling concentration is also $\langle n \rangle = 0.933$ and $\langle n \rangle = 0.893$. It can be clearly observed from Fig. 9 that with the synchronous decrease of the strength of the electron hopping breaking term $t'_{1,2}$, for the case of electron filling concentration $\langle n \rangle = 0.933$, except for the average effective pairing correlation function strength of the triplet chiral NN-p+ip and NNN-p+ip wave pairing symmetry, which has no obvious change, and the singlet NN-s wave, which shows a weak decrease, the singlet chiral NN-d + id and NNN-d+id and the triplet NNN- f wave pairing symmetry. In addition, for the chiral NN-d+id wave, the specific increase of the average effective pairing correlation function is based on the average effective pairing correlation function strength at $t'_1 = t'_2 = 0.10$, which increases by 23. 15% at $t'_1 = t'_2 = 0.05$ and decreases by 11. 22% at $t'_1 = t'_2 = 0.15$. For the case of electron filling concentration $\langle n \rangle = 0.893$, the difference is that the strength of the average effective pairing correlation function of the singlet NN-s wave pairing symmetry does not change significantly, and the strength of the singlet NNN-d + id wave even decreases due to the influence of calculation errors. The same point is that the change trend of other types of electron pairing symmetry is almost the same, and the singlet chiral NN-d+id wave still occupies a dominant position. Similarly, based on the average effective pairing correlation function strength at $t'_1 = t'_2 = 0.10$, the average effective pairing correlation function strength of the chiral NN-d+id wave of the singlet state at $t'_1 = t'_2 = 0.05$ increases by 18.99%; The strength of the average effective pairing correlation function decreases by 93.98% at $t'_1 = t'_2 = 0.15$. Through the above analysis, it can be predicted that the superconducting critical transition temperature of bilayer graphene will be significantly increased when the twist angle is around 1.08 ° and gradually decreases, that is, the system will have more robust superconductivity at a smaller twist angle. Secondly, the system still has robust superconductivity when the twist angle is greater than 1.08 ° at high electron filling concentration $\langle n \rangle = 0.933$; However, when the twist angle is greater than 1.08 ° at low electron filling $\langle n \rangle = 0.893$, the superconducting state of the system will be destroyed due to the sharp decrease of the average effective pairing correlation function of the chiral NN-d+id wave.

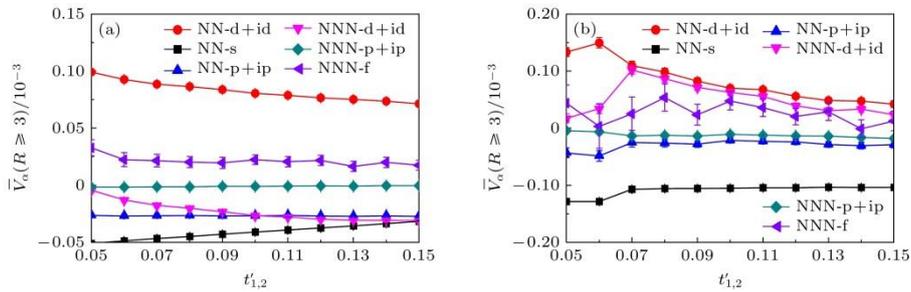

**Figure 9.** Average effective pairing correlation functions $\overline{V}\alpha(R \geqslant 3)$ as a function of electron hopping anisotropy terms $t'_{1,2}$ for various electron pairing symmetries in a system with on-site Coulomb interaction strength $U = 2.0$, nearest-neighbor Coulomb interaction strength $V = 0.0$, and lattice size $L = 5$: (a) $\langle n \rangle = 0.933$; (b) $\langle n \rangle = 0.893$.

In order to further verify the reliability of the above calculation results, the numerical simulation is further extended to the lattice scale of $L = 6$, and the evolution curve of the effective pairing correlation function of chiral NN-d+id wave pairing symmetry under the control of flat band structure on the long-range distance $R \geqslant 3$ is shown Fig. 10. The calculation parameters selected in this figure are set to the nearest neighbor Coulomb interaction strength $= 0.0$, the on-site Coulomb interaction strength and the electron filling concentration are consistent with Fig. 6. As shown in the Fig. 10, it can be observed that the strength of the effective pairing correlation function of the chiral NN-d+id wave pairing symmetry is significantly enhanced over the whole long range when the energy band is narrower and flatter by adjusting the electron hopping breaking term. In addition, it is noted that the strength of the effective pairing correlation function of the chiral NN-d+id wave pairing symmetry over the whole long range tends to decrease when the band structure is made broader and more convex by adjusting the electron hopping breaking term. Therefore, the results confirm that when the twist angle deviates from 1.08 °, the further decrease of the twist angle is beneficial to the enhancement of superconductivity, while the increase of the twist angle will inhibit the superconductivity of the system.

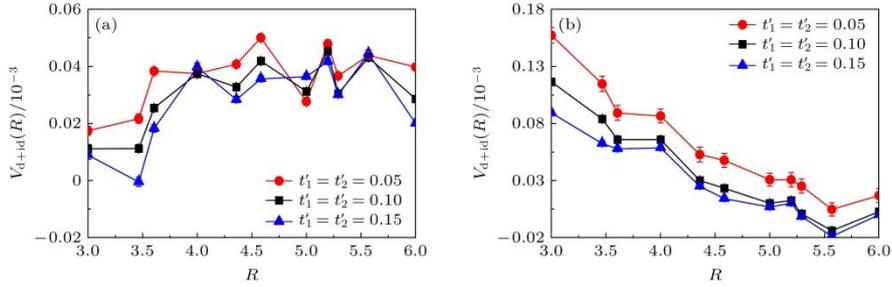

**Figure 10. Effective pairing correlation function $V_{d+id}$ as a function of the long-range pairing distance $R$ for the chiral NN-$d + id$-wave pairing symmetry with lattice size $L = 6$ under flat band structure modulation with an on-site Coulomb interaction strength of $U = 2.0$ and nearest-neighbor Coulomb interaction strength of $V = 0.0$: (a) $\langle n \rangle = 0.954$; (b) $\langle n \rangle = 0.926$.**

In the previous discussion, it was mentioned that magnetism and superconductivity are closely related. Next, we will focus on the relationship between magnetism and superconductivity under flat-band control. The Fig. 11 shows the variation of the spin structure factor $S(q)$ along the direction of the high-symmetry line $\Gamma \to M \to K \to \Gamma$ in the first Brillouin zone. The calculated parameters selected in this figure are consistent with the Fig. 7 except for the fixed nearest neighbor Coulomb interaction strength $V = 0.0$. First of all, based on this result, it can be clearly observed that the spin structure factor has the largest value near the $\Gamma$ point, which indicates that the system exhibits robust antiferromagnetic order, and indirectly indicates that the electron pairing form of the twisted bilayer graphene superconductor is the chiral NN-d+id wave pairing symmetry of the singlet state. Secondly, in the process of controlling the flat band structure of the twisted bilayer graphene by adjusting the electron hopping breaking term, when the electron hopping breaking term is $t'_1 = t'_2 = 0.05$, the narrow and flat band structure will induce the spin structure factor to show an increasing trend near the $\Gamma$ point; When the electron hopping term $t'_1 = t'_2 = 0.15$ is broken, the wide and convex band structure will lead to a weakening trend of the spin structure factor near the $\Gamma$ point, which further indicates that the control of the flat band structure to make the system show a robust antiferromagnetic order is an important condition for the formation of chiral NN-d+id superconducting state.

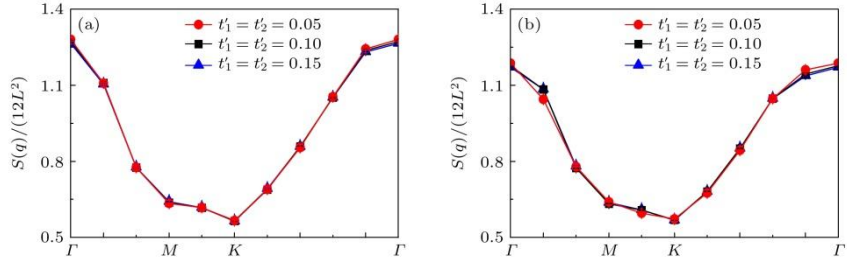

Figure 11. **Spin structure factor** $S(q)$ **along the high-symmetry lines** $\Gamma \to M \to K \to \Gamma$ **in the first Brillouin zone for a system with on-site Coulomb interaction strength** $U = 2.0$, **nearest-neighbor Coulomb interaction strength** $V = 0.0$, **and lattice size** $L = 5$, **under flat band structure modulation: (a)** $\langle n \rangle = 0.933$; **(b)** $\langle n \rangle = 0.893$.

## 4. Conclusion

In this paper, we use quantum Monte Carlo method to study the intrinsic relationship between magnetism and superconductivity in the effective two-orbital Hubbard model of twisted bilayer graphene, and explore the methods to improve the superconducting critical transition temperature of twisted bilayer graphene system by numerical simulation. Firstly, the control of superconductivity by the attractive nearest-neighbor Coulomb interaction is studied to illustrate the important influence of electron-phonon coupling on the torsional bilayer graphene system. The numerical results not only show that the superconducting state of the twisted bilayer graphene system is the chiral $\text{NN-d} + \text{id}$ wave dominated electron pairing form, but also show that the nearest neighbor attractive Coulomb interaction can greatly induce the increase of the effective pairing correlation function of the chiral $\text{NN-d} + \text{id}$ wave over the whole long-range distance. So far, it is confirmed that the attractive nearest neighbor Coulomb interaction has a positive effect on improving the superconductivity of the system.

Secondly, the effect of the twist angle on the superconducting state is studied, and the flat band structure is constructed based on the electron hopping breaking term to reflect different twist angles. The numerical simulation results show that the effective pairing correlation function of chiral NN-d + id wave has a large increase when the twist angle deviates from 1. 08 ° and decreases gradually, while the effective pairing correlation function of chiral NN-d + id wave has a decreasing trend when the twist angle increases. This result predicts that the torsion bilayer graphene system may have a higher superconducting critical transition temperature when the torsion angle is further reduced. Finally, the reason for the modulation of superconductivity by the attractive neighbor Coulomb interaction and the flat band structure is analyzed from the magnetic point of view. Based on the enhancement of the spin structure factor near the$\Gamma$ point in the Brillouin zone, it is shown that the enhancement of the antiferromagnetic order is a prerequisite for the increase of the superconducting critical transition temperature of the system, and is also a favorable condition for the formation of chiral NN-d + id wave. Through this series of studies, the results not only help to understand the strongly correlated systems such as torsion bilayer graphene more comprehensively, but also provide a research direction for finding torsion systems with higher superconducting critical transition temperature.

I would like to thank Professor Zhang Yan of Northwest Normal University and Dr. Yang Hui of Hubei Second Normal University for their active discussions.